\documentclass[sigconf]{acmart}

\usepackage{balance}

\def\BibTeX{{\rm B\kern-.05em{\sc i\kern-.025em b}\kern-.08emT\kern-.1667em\lower.7ex\hbox{E}\kern-.125emX}}

\usepackage{url}
\usepackage{graphicx}
\usepackage{amsmath,amssymb,amsfonts}
\usepackage{algorithmic}
\usepackage{textcomp}
\usepackage{xcolor}
\usepackage{tikz}
\usetikzlibrary{topaths,calc}
\usepackage[ruled]{algorithm2e}
\usepackage{subfigure}
\usepackage{booktabs}
\usepackage{balance}

\newtheorem{myDef}{Definition}

\copyrightyear{2019}
\acmYear{2019} 
\setcopyright{acmcopyright}
\acmConference[CIKM '19] {The 28th ACM International Conference on Information and Knowledge Management}{November 3--7, 2019}{Beijing, China}
\acmBooktitle{The 28th ACM International Conference on Information and Knowledge Management (CIKM'19), November 3--7, 2019, Beijing, China}
\acmPrice{15.00}
\acmDOI{10.1145/3357384.3357871}
\acmISBN{978-1-4503-6976-3/19/11}



\begin{document}

\fancyhead{}

\title{Hyper-Path-Based Representation Learning for Hyper-Networks}

\author{Jie Huang}
\affiliation{\institution{Sun Yat-sen University}}
\email{huangj285@mail2.sysu.edu.cn}

\author{Xin Liu}
\affiliation{\institution{The Hong Kong University of Science and Technology}}
\email{xliucr@cse.ust.hk}

\author{Yangqiu Song}
\affiliation{\institution{The Hong Kong University of Science and Technology}}
\email{yqsong@cse.ust.hk}

\begin{abstract}

Network representation learning has aroused widespread interests in recent years. While most of the existing methods deal with edges as pairwise relationships, only a few studies have been proposed for hyper-networks to capture more complicated tuplewise relationships among multiple nodes. A hyper-network is a network where each edge, called hyperedge, connects an arbitrary number of nodes. Different from conventional networks, hyper-networks have certain degrees of indecomposability such that the nodes in a subset of a hyperedge may not possess a strong relationship. That is the main reason why traditional algorithms fail in learning representations in hyper-networks by simply decomposing hyperedges into pairwise relationships. In this paper, we firstly define a metric to depict the degrees of indecomposability for hyper-networks. Then we propose a new concept called hyper-path and design hyper-path-based random walks to preserve the structural information of hyper-networks according to the analysis of the indecomposability. Then a carefully designed algorithm, Hyper-gram, utilizes these random walks to capture both pairwise relationships and tuplewise relationships in the whole hyper-networks. Finally, we conduct extensive experiments on several real-world datasets covering the tasks of link prediction and hyper-network reconstruction, and results demonstrate the rationality, validity, and effectiveness of our methods compared with those existing state-of-the-art models designed for conventional networks or hyper-networks.

\end{abstract}


\maketitle

\section{Introduction \label{sec:intro}}

\begin{figure}[tp!]
\centerline{\includegraphics[width=0.9\linewidth]{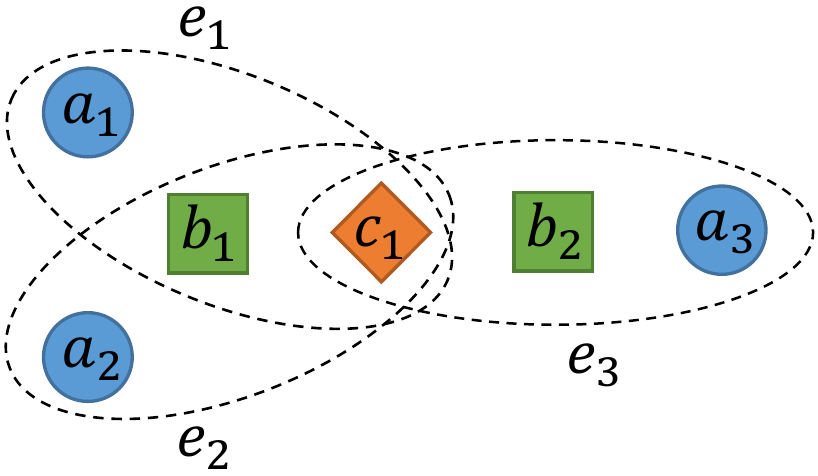}}
\caption{An illustration of a $3$-uniform heterogeneous hyper-network. In this hyper-network, there are three types of nodes (e.g., $\phi(a_1)=a, \phi(b_1)=b, \phi(c_1)=c$) and three hyperedges ($e_1 = \{a_1,b_1,c_1\},e_2 = \{a_2,b_1,c_1\},e_3 = \{a_3,b_2,c_1\}$).}
\label{fig:hyper-network}
\end{figure}

Networks or graphs are data structures to model the relationships between objects in a system. There are a large number of networks presenting different types of relationships in various domains, and examples include social networks, biological networks, brain networks, etc. Due to the importance of networks, many methods are proposed for network analysis tasks such as link prediction \cite{liben2007link} and similarity search \cite{jeh2003scaling,sun2011pathsim}. Among these methods, network representation learning methods \cite{cui2018survey,cai2018comprehensive,perozzi2014deepwalk,tang2015line,grover2016node2vec} which aim to learn the low-dimensional representations of nodes in networks have aroused widespread interest in recent years. 
Most of the existing network representation learning methods are designed for conventional networks, where each edge represents a pairwise relationship between two nodes.
However, relationships among objects are much more complex in the real world.
In particular, it is hard to represent such a connection (user, drug, reaction) in a conventional network, which corresponds to a record that a user produces a specific side effect after taking a kind of drug.
In this case, it is more reasonable to consider the interactions among these objects as tuplewise relations, where each tuplewise relation involves more than two nodes.
Such complex tuplewise relations are usually modeled as hyperedges which indicate relationships among an arbitrary number of nodes, and such systems are modeled as hyper-networks (refer to Fig. \ref{fig:hyper-network} as an example).

A typical way to analyze hyper-networks is to decompose hyperedges into pairwise edges by clique expansion \cite{sun2008hypergraph} or star expansion \cite{agarwal2006higher}, and then apply traditional network representation learning methods. This way assumes that hyperedges are decomposable.
However, there is no theory or proof to support the correctness.
\citeauthor{tu2018structural} \cite{tu2018structural} investigate the indecomposability of heterogeneous hyper-network: the indecomposability of hyper-networks means that a set of nodes in a hyperedge has a strong tuplewise relationship, while the nodes in a subset of the hyperedge do not necessarily have a strong relationship. Besides, they prove that for an indecomposable hyper-network, any linear tuplewise similarity function cannot satisfy the indecomposable property.
Their work implies the analysis of hyper-networks should take the indecomposability into consideration, but they simply agree that homogeneous hyper-networks are usually decomposable and heterogeneous ones are indecomposable.
This trend is true in some cases, but not for all cases. 
There is no doubt that some negative bias will be introduced in algorithms if we simply treat hyper-networks as indecomposable or decomposable ones.

Different hyper-networks have different degrees of indecomposability. For a (user, drug, reaction) hyper-network, 
it is not convincing to establish a relationship between a user and a drug reaction without information about a kind of drug, and it is also meaningless to connect a user and a kind of drug without a drug reaction.
However, considering a (user, movie, tag) hyper-network, relations between movies and tags can appear independently apart from users 
if we assume the tags of one movie are mainly determined by its content,
relations between users and movies can be separated from tags 
because each user has a preference for a specific movie type,
and relations between users and tags can also be independent 
due to user habits.
Besides, the degrees of indecomposability of different types of subsets in a hyper-network may vary.
For instance, the (user, drug) and (user, reaction) pairs cannot be independent in the (user, drug, reaction) hyper-network described above, but drugs and reactions do have some relationships because a kind of drug usually has some common side effects. Under the circumstance, (drug, reaction) can be independent to a certain degree.

On the other hand, network representation learning methods based on random walks play an important role in the community. Inspired by Word2vec \cite{mikolov2013distributed}, DeepWalk \cite{perozzi2014deepwalk} firstly treats the random walk sequences in networks as the sentences in documents and learns the representations for nodes in a similar way of learning representations for words. 
Node2vec \cite{grover2016node2vec} is then proposed to make random walks more flexible by introducing two parameters. 
Additionally, there are some random walk-based methods designed for networks with special structures. Among them, BiNE \cite{gao2018bine} is proposed for bipartite network representation learning and Metapath2vec \cite{dong2017metapath2vec} is proposed for heterogeneous network representation learning. 
However, these existing random walk-based methods do not take advantage of the indecomposability of the hyper-networks thus cannot generate random walks that well preserve the structural information of hyper-networks.

Based on the above considerations, we firstly define a metric, named indecomposable factor, 
to depict the degrees of indecomposability for hyper-networks.
Especially for uniform heterogeneous hyper-networks, our metric is a good measurement for each node type of indecomposability. 
In order to preserve such complex information about the indecomposability,
we further introduce a new concept called hyper-path that we select the next node according to some special rules on each step of the random walks. 
With a good assessment of the indecomposability of hyper-networks, we design a flexible hyper-path-based random walk model to generate random walks for hyper-networks with different degrees of indecomposability.
Finally, we propose a deep model called Hyper-gram to capture both pairwise and tuplewise relationships in the hyper-path-based random walks, and low-dimensional representations and a nonlinear tuplewise similarity function are produced at the same time.

Extensive experiments on several real-world hyper-networks with different degrees of indecomposability are conducted. There are two well-known tasks in our experiments: link prediction, where we predict the unobserved hyperedges of the original hyper-network, and hyper-network reconstruction, where we reconstruct the hyper-network with the low-dimensional vector representations of nodes. Experimental results demonstrate that our proposed model can well mine the hidden tuplewise relationships of original hyper-networks and reconstruct hyper-networks with a minimum error compared with the existing state-of-the-art methods.

Our main contributions are as follows:
\begin{itemize}
    \item We define a metric called \textbf{indecomposable factor} to depict the degrees of indecomposability for hyper-networks and apply it on several real-world hyper-networks.
    \item We introduce a new concept called \textbf{hyper-path} and design a flexible hyper-path-based random walk model to preserve the complex information about indecomposability and local structure for hyper-networks.
    \item We propose a deep model called \textbf{Hyper-gram} to learn the representations of nodes in hyper-networks which preserves both pairwise similarity and tuplewise similarity of hyper-networks meanwhile training a tuplewise similarity function to evaluate the tuplewise relationships among nodes. 
    \item We conduct experiments on four real-world hyper-networks in different domains. Results show that our model achieves the state-of-the-art performance in two important tasks compared with network representation learning methods on conventional networks and hyper-networks.
\end{itemize}

The remainder of this paper is organized as follows. In Section 2, we review the related work. In Section 3, we delve into details of the proposed methods. In Section 4, we perform extensive empirical studies. We conclude our work in Section 5.
The code and data are available at \url{https://github.com/HKUST-KnowComp/HPHG/}.

\section{Related Work}

Our work is related to network representation learning methods on conventional networks. Some classical networks representation learning methods are based on matrix factorization, such as Isomap \cite{tenenbaum2000a}, Local Linear Embedding \cite{roweis2000nonlinear} and Laplacian Eigenmaps \cite{belkin2002laplacian}. There are two main problems in these methods. First, these methods are hard to apply to large-scale networks due to the computationally expensive eigendecomposition operations \cite{wang2016scalable,wang2017learning}. Second, these models are not flexible and are sensitive to the predefined proximity measures which make strong assumptions about the relationship between the network structure and the prediction task.
Recently, DeepWalk \cite{perozzi2014deepwalk} and Node2vec \cite{grover2016node2vec} introduce random walk model to expand the neighborhood of nodes of networks with the help of Word2vec \cite{mikolov2013distributed}.
Besides, LINE \cite{tang2015line} designs and optimizes the objective function to preserve both the first order and second order proximities of networks. Beyond first order and second order proximities, AROPE \cite{zhang2018arbitrary} is a model that supports shifts across arbitrary order proximities based on SVD framework. In addition, there are some deep neural network based methods such as SDNE \cite{wang2016structural}, SDAE \cite{cao2016deep}, and SiNE \cite{wang2017signed}, which introduce deep models to fit network data.
The methods described above are mainly designed for homogeneous networks where the types of nodes are the same.
There are also some methods such as HNE \cite{chang2015heterogeneous}, EOE \cite{xu2017embedding}, and Metapath2vec \cite{dong2017metapath2vec} proposed for heterogeneous network representation learning.

However, these network representation learning methods only consider the pairwise relationships between nodes in conventional networks. 
In the real world, some problems are required to consider the complex interactions among more than two nodes, where hyper-networks are used to model such tuplewise relationships. In fact, hyper-networks are common in the real world, but only several existing representation learning methods are designed for them. There are some methods \cite{ng2002spectral,zhou2007learning,wu2010multiple} based on spectral clustering, but these methods make a strong homophily assumption that graph cuts will be useful for classification \cite{tang2011leveraging}. Such assumptions are not reasonable in all scenarios so that the performance is not satisfactory in most cases. There are also several methods proposed recently for hyper-network representation learning. Among them, DHNE \cite{tu2018structural} is a deep model to learn a nonlinear tuplewise similarity function for link prediction, where the authors take the indecomposability of hyperedges into account. With a similar idea, HHNE \cite{baytas2018heterogeneous} is a deep model based on the graph convolutional network, which integrates the features of nodes.
Hyper2vec \cite{huang2019hyper2vec} introduce a flexible random walk model on hyper-networks in a semi-supervised setting.
HHE \cite{zhu2016heterogeneous} utilizes eigenvectors to find the optimal solution and learn the subspace representations, but the computational cost is a vital problem.
HGE \cite{yu2018modeling} designs an objective function for hyperedges to make involved node representations close to each other, but the model is not flexible for hyper-networks with different structures. Besides, both HHE and HGE do not combine a tuplewise similarity function to evaluate the tuplewise relationships of nodes.

\section{Methodlogy}

In this section, we introduce the background of hyper-network representation learning at the beginning. For our methods, we first define the indecomposable factor to evaluate the degrees of indecomposability for hyper-networks, then introduce the definitions of the hyper-path and the hyper-path-based random walk, and finally propose the Hyper-gram model which learns the final representations of nodes by integrating the pairwise and tuplewise similarities contained in the generated hyper-path-based random walks.

\subsection{Background}

\paragraph{Hyper-network}

A \textbf{hyper-network} is defined as a hypergraph $\mathcal{G}(\mathcal{V}, \mathcal{E})$ with a node type mapping function $\phi: \mathcal{V} \to \mathcal{A}$, where each node $v \in \mathcal{V}$ belongs to one particular node type $\phi(v) \in \mathcal{A}$. A \textbf{hyperedge} $e \in \mathcal{E} $ which is a subset of $\mathcal{V}$ indicates the tuplewise relationship among an arbitrary number of nodes. 
The neighbors of a node $v$ is a node set $\mathcal{N}_\mathcal{G}(v) = \{u|\exists e \in \mathcal{E} \quad (v \in e) \land (u \in e)\}$. Given a set $\mathcal{S}$, $|\mathcal{S}|$ denotes the cardinality of $\mathcal{S}$. In particular, the hyper-network is called homogeneous when $|\mathcal{A}| = 1$ and heterogeneous when $|\mathcal{A}| > 1$. And a hyper-network is called uniform when all the hyperedges contain the same number of nodes. An illustration of a $3$-uniform heterogeneous hyper-network is shown in Fig. \ref{fig:hyper-network}.

\paragraph{Hyper-network Representation Learning} Hyper-network representation learning aims to learn a low-dimensional representation for each node in the hyper-network while preserving the global and local network structures. In this paper, the objective of hyper-network representation learning is to learn a node vector mapping function $f: \mathcal{V} \to X \in \mathbb{R}^{|\mathcal{V}|\times d}$ and a tuplewise similarity function $s_{tuple}: \mathcal{T} \to s \in [0,1]$, where $d$ is the dimension of the low-dimensional vectors and $\mathcal{T}$ is the set of possible tuples containing several nodes in $\mathcal{V}$.

\subsection{Hyper-Network Indecomposability \label{sec:indecomposability}}

The most important property of hyper-network is the indecomposability of hyperedges, which is different from the conventional network in which there are only pairwise relationships. For an indecomposable hyperedge which represents a tuplewise relationship of a set of nodes, the nodes in the subset of the hyperedge do not necessarily have a strong relationship. 

However, as discussed above, different hyper-networks have different degrees of indecomposability, and even the indecomposability of different subsets of a hyperedge is also different. 
In order to make the model suitable for hyper-networks with different degrees of indecomposability, a metric to evaluate the degrees of indecomposability for hyper-networks is necessary.

The metric we proposed is based on the statistical analysis of hyperedges. For a specific hyperedge $e$, there are two events related to $e$:
\begin{itemize}
    \item event $A$: $e \in \mathcal{E}$;
    \item event $B_t$: $\exists e'\in \mathcal{E}, \exists v \in e \quad (\phi(v) = t) \land (e'\neq e) \land (e-\{v\}\subseteq e')$,
\end{itemize}
\noindent where $e-\{v\}$ is the subset of $e$ without $v$. Given a hyperedge $e$, event $B_t$ indicates whether there exists another hyperedge $e'$ containing a particular subset of it. Given the hyper-network illustrated in Fig. \ref{fig:hyper-network} as an example, for $e_1 = \{a_1,b_1,c_1\}$, event $A$ is true because $e_1 \in \mathcal{E}$ and event $B_a$ is true because $\exists e_2 \in \mathcal{E}, \exists a_1 \in e_1 \quad (\phi(a_1) = a) \land (e_2\neq e_1) \land ( e_1-\{a_1\} = \{b_1,c_1\}\subseteq e_2)$. However, event $B_b$ and event $B_c$ are not true because there are no other hyperedges that meet the requirements. After a good understanding of event $A$ and event $B_t$, we define the indecomposable factor $\xi_t$ as follows:
\begin{myDef}{(Indecomposable Factor)}
\begin{equation}
\xi_t = \frac{p(B_t)}{p(B_t|A)}.
\end{equation}
\end{myDef}

The idea of the indecomposable factor is based on the fact that for a hyperedge $e$, event $A$ means that there is a tuplewise relationship among the nodes in $e$. If $e$ is decomposable, the nodes in the subset of $e$ also possess some relationships.  Therefore, when event $A$ is true, event $B_t$ is more likely to be true, which means that $p(B_t|A)$ is larger than $p(B_t)$. On the other hand, if $e$ is indecomposable, which means that Event $A$ and Event $B_t$ are not associated, thus we have $p(B_t|A) \approx p(B_t)$. In general, the larger the factor, the stronger the indecomposability of the hyper-network, and if a hyper-network is randomly generated, the value of the indecomposable factor is close to 1.

To calculate the indecomposable factor, we need to calculate $p(B_t)$ and $p(B_t|A)$ respectively. Approximately, we calculate $\xi_t$ as follows:
\begin{equation} 
\xi_t = \frac{{\frac{1}{|\mathcal{E}_{random}|} \sum_{e \in \mathcal{E}_{random}} \delta_1(e)}}{\frac{1}{|\mathcal{E}|} \sum_{e \in \mathcal{E}} \delta_1(e)},
\end{equation}
\noindent where $\mathcal{E}_{random}$ is an edge set generated randomly from the node set and the distribution of edge degree according to the distribution of edge degree in $\mathcal{E}$.
$|\mathcal{E}_{random}|$ depends on the number of edges, e.g., $10 \times |\mathcal{E}|$, and we can sample more random edges to make the results more accurate for small hyper-networks. $\delta_1(\cdot)$ is an indicator function defined as follows:
\begin{equation}
\delta_1(e) = 
\begin{cases}
1, \quad \text{event $B_t$ is true} \\
0, \quad \text{otherwise}
\end{cases}.
\end{equation}

\subsection{Hyper-Path-Based Random Walk \label{sec:hyper-path}}

The traditional random walk models which simply select the next node from the neighbors of the current node at random or based on the previous transition probability described in \cite{zhou2007learning} cannot capture the indecomposability of hyper-networks. Using the hyper-network illustrated in Fig. \ref{fig:hyper-network} as an example, we assume that the hyperedges are highly indecomposable, which implies that without type $b$ nodes, there is no strong correlation between a type $a$ node and a type $c$ node. In this case, the similarity between $a_1$ and $a_2$ are much higher than the similarity between $a_1$ and $a_3$. However, the traditional random walk models also give a high probability for $a_1$ to include $a_3$ in its context (e.g., path $a_1$-$c_1$-$b_2$-$a_3$), and the meta-path-based model also generates paths like $a_1$-$b_1$-$c_1$-$b_2$-$a_3$ to close the relationship between $a_1$ and $a_3$. Thus none of the traditional models can capture the indecomposability of hypergraphs well. To this end, a new concept called \textit{hyper-path} is introduced in our work. Before introducing the hyper-path, we first talk about the \textit{Path Order (PO)} which is defined as follows:
\begin{myDef}{(Path Order)} 
For a path $P$, the \textbf{path order} of node $v$ is calculated as follows:
\begin{align}
& PO(v|P) = \max k \notag \\
\text{s.t.~} & 1. \exists e\in \mathcal{E} \quad (P[-k],\cdots,P[-1]\in e) \land (v \in e) \notag \\
& 2. P[-k] \neq \dots \neq P[-1] \neq v,
\end{align}
\noindent where $P[-k]$ is the last $k$-th element of path $P$.
\end{myDef}

Given a node and a path, the path order of the node depends on whether there is a hyperedge containing the node and the last $k$ different nodes of the path at the same time. Take Fig. \ref{fig:hyper-network} as an example, when path $P = \text{$a_1$-$b_1$-$c_1$}$, we have $PO(a_2|P)=2$ and $PO(a_3|P)=PO(b_2|P)=1$. With the definition of path order, our proposed hyper-path is defined as follows:
\begin{myDef}{(Hyper-path)}
A \textbf{hyper-path} is a path which is generated based on the following rule: starting from node $v_s$, select a node from the nodes with the largest path order as the next node of the current path.
\end{myDef}

Hyper-paths can well preserve the tuplewise relationships of nodes in hyper-networks. Furthermore, hyper-paths can also better enhance the pairwise relationships between nodes compared to traditional paths. 
Take Fig. \ref{fig:hyper-network} as an example, 
four hyper-paths with the same start point and length,
$a_1$-$b_1$-$c_1$-$a_1$, $a_1$-$b_1$-$c_1$-$a_2$, $a_1$-$c_1$-$b_1$-$a_1$ and $a_1$-$c_1$-$b_1$-$a_2$, 
are able to imply the strong pairwise relationship between $a_1$ and $a_2$. But they do not support the correlation between $a_1$ and $a_3$. 
Additionally, sub-paths such as $a_1$-$b_1$-$c_1$ and $b_1$-$c_1$-$a_2$ can provide signals of tuplewise relationships of nodes in the hyper-network with high confidence.

As discussed above, different hyper-networks have different degrees of indecomposability, and different types of subsets of a hyperedge also have different indecomposable factors. On one hand, we need to generate random walks with the property of hyper-path to preserve the indecomposability of hyper-networks to some extent; on the other hand, the random walker should have some exploration ability to mine more potential relationships.
Based on the above-mentioned proposes,
we use the indecomposable factor defined before as a measurement to guide random walks. The unnormalized transition probability of the \textit{hyper-path-based random walks} is calculated as follows:
\begin{equation}
\pi_2(v|P) = \pi_1(v|P[-1]) \cdot \exp(\alpha \cdot \xi_{\phi(v)} \cdot (PO(v|P)-1)),
\end{equation}
\noindent where $\alpha$ is a coefficient to control the tendency toward hyper-path and $\pi_1(v|P[-1])$ is the first order transition probability which can be simply defined as follows:
\begin{equation}
\pi_1(v|P[-1]) = 
\begin{cases}
1, \quad v \in \mathcal{N}_\mathcal{G}(P[-1]) \\
0, \quad \text{otherwise}
\end{cases}.
\end{equation}

\noindent Based on the transition probability, if the indecomposable factor is large, the generated random walks are close to hyper-paths. Otherwise, the generated random walks will not have too much bias, which is close to the traditional random walks. 

To explain the principle of the hyper-path-based random walks, we use a $3$-uniform hyper-network with types (a, b, c) as an example. 
Assuming $\phi(P[-2])=a$ and $\phi(P[-1])=b$, if $\xi_c$ is large, which means that the relationship between a type $a$ node and a type $b$ node cannot exist independently of the type $c$ nodes, thus we select the next nodes which possess a higher path order with a higher probability; if $\xi_c$ is small, which means that the relationship between a type $a$ node and a type $b$ node can exist independently of the type $c$ nodes, thus we do not give too much priority to the type $c$ nodes with a higher path order, but let the random walker explore more possible neighbors.

\subsection{Hyper-Gram Model \label{sec:hyper-gram}}

\begin{figure*}[tp!]
\centerline{\includegraphics[width=0.97\linewidth]{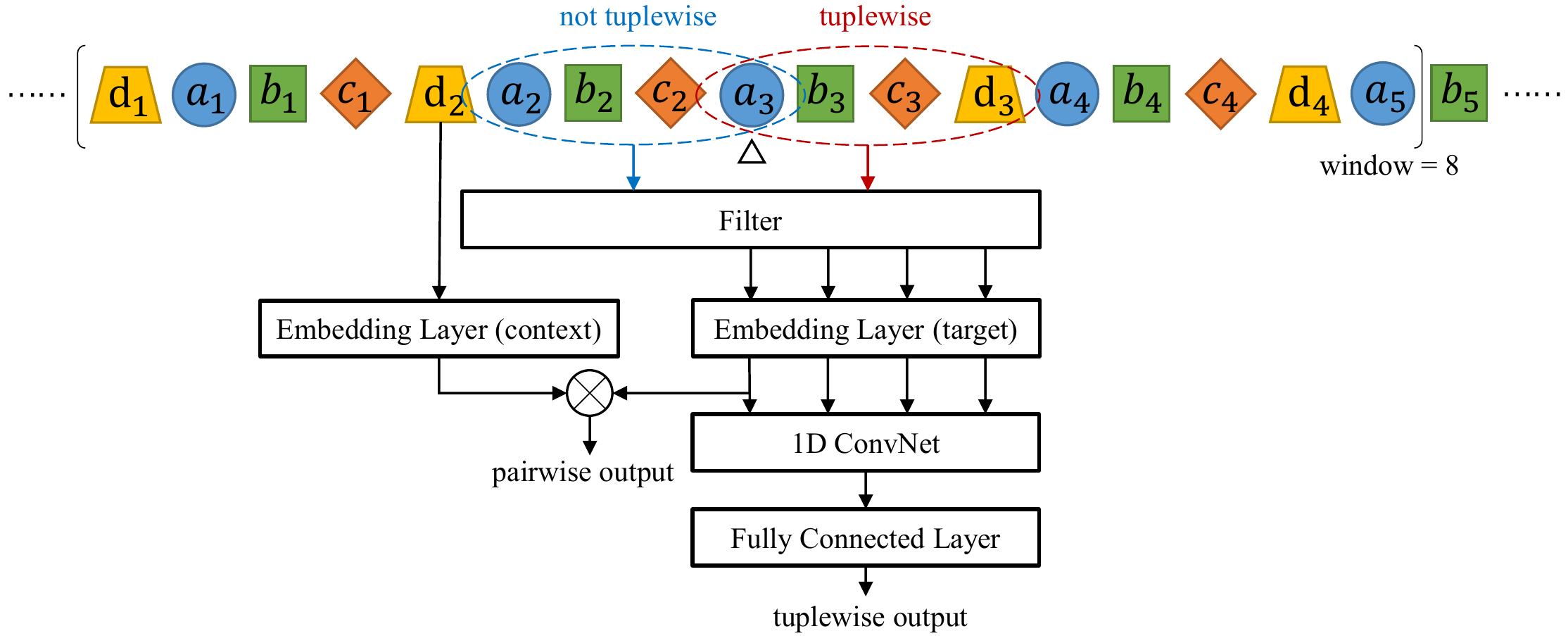}}
\caption{An illustration of the Hyper-gram model. In this case, the hyper-network is $4$-uniform and the window size is 8. 
There are two possible tuples $(a_2,b_2,c_2,a_3)$ and $(a_3,b_3,c_3,d_4)$ possessing tuplewise relationships, but the former is not in line with the property of the hyper-network (there is no node with type $d$), so the model filters it out.}
\label{fig:Hyper-gram}
\end{figure*}

For hyper-networks with large indecomposable factors, hyper-path-based random walks have advantages as follows:

\begin{itemize}
\item For two nodes close to each other in the hyper-path-based random walks, the pairwise relationship of these two nodes is stronger than that in the traditional random walks.
\item Several consecutive nodes in the hyper-path-based random walks may possess a strong tuplewise relationship.
\end{itemize}

In order to learn the representations for nodes from random walks, a traditional way is to use the Skip-gram model.
Skip-gram \cite{mikolov2013distributed} is originally designed for language modeling. It learns low-dimensional distributed representations of words by maximizing the pairwise co-occurrence probabilities among words that appear within a window in a sentence. Inspired by the Skip-gram model, DeepWalk \cite{perozzi2014deepwalk} replaces sentences in documents with random walks in networks to learn representations of nodes in networks efficiently.
Skip-gram can well capture the pairwise relationships in the traditional random walks, but it is not able to gain from the tuplewise relationships in our designed hyper-path-based random walks. 
Besides, in order to evaluate the tuplewise relationship of several nodes, a nonlinear tuplewise similarity function is necessary.

Based on the above considerations, we propose a deep model called Hyper-gram to learn the representations for nodes in hyper-networks based on the hyper-path-based random walks. The framework of Hyper-gram is shown in Fig. \ref{fig:Hyper-gram}.

Unlike the Skip-gram model which only considers pairwise relationships, 
Hyper-gram tries to maximize the pairwise similarity and the tuplewise similarity jointly.

\paragraph{Pairwise Similarity}

Inspired by DeepWalk and Node2vec, we want to combine the idea of Skip-gram and hyper-path-based random walks to optimize the pairwise similarity between two nodes.
Skip-gram aims to maximize the log-probability of observing context words for a center word. 
Since random walks are generated based on the connectivity structure among nodes in conventional networks or hyper-networks, 
we can regard random walks as sentences and maximize the log-probability of observing nodes in a context window for a center node.
Given a specific node $u$ and a specific node $v$, we want to calculate $p(u|v)$, which is the probability that $u$ is a context node for $v$.
In Skip-gram, the conditional probability distribution is given by taking vector dot-products and applying
the softmax function:
\begin{equation}
    p(u|v) = \frac{\exp(f(v) \cdot f'(u))}{\sum_{w \in \mathcal{V}} \exp(f(v) \cdot f'(w))},
    \label{eq:conditional_distribution}
\end{equation}
\noindent where $f(\cdot)$ is the center embedding layer and $f'(\cdot)$ is the context embedding layer, $\mathcal{V}$ is the set of nodes in the hyper-networks.
We can further define a pairwise similarity function between two nodes $v_1, v_2$ based on Eq.~\eqref{eq:conditional_distribution} as follows:
\begin{equation}
    s_{pair}(v_1, v_2) = \prod_{w \in \mathcal{V}}{p(w|v_1)p(w|v_2)}.
\end{equation}
\noindent $s_{pair}(v_1, v_2)$ is large when $v_1$ and $v_2$ share similar context nodes.
The logarithm of $s_{pair}(v_1, v_2)$ is easier to optimize:
\begin{align}
    &\log{s_{pair}(v_1, v_2)} \nonumber 
    = \sum_{w \in \mathcal{V}}{\Big(\log{p(w|v_1)} + \log{p(w|v_2)}\Big)} \nonumber \\
    & \ \ = \sum_{w \in \mathcal{V}}{\Big( f(v_1) \cdot f'(w) - \log{Z_{v_1}} + f(v_2) \cdot f'(w) - \log{Z_{v_2}} \Big)},
    \label{eq:log_s_pair}
\end{align}
\noindent where $Z_{v_1}=\sum_{w \in \mathcal{V}} \exp(f(v_1) \cdot f'(w))$ and $Z_{v_2}=\sum_{w \in \mathcal{V}} \exp(f(v_2) \cdot f'(w))$.
However, $Z_{v_1}$ and $Z_{v_2}$ are expensive to compute, and we approximate Eq. \eqref{eq:log_s_pair} using negative sampling~\cite{mikolov2013distributed} and add an activation function:
\begin{align}
    \sum_{w \in \mathcal{V}} & \Big(\log{\sigma(f(v_1) \cdot f'(w))} - \sum_{w_{n1} \in \mathcal{V}_{n1}}{\log{\sigma(f(v_1) \cdot f'(w_{n1}))}} \nonumber\\
    & + \log{\sigma(f(v_2) \cdot f'(w))} - \sum_{w_{n2} \in \mathcal{V}_{n2}}{\log{\sigma(f(v_2) \cdot f'(w_{n2}))}} \Big),
\end{align}
\noindent where $\sigma(\cdot)$ is the sigmoid function, $\mathcal{V}_{n1}$ is the set of negative samples for $v_1$, $\mathcal{V}_{n2}$ is the set of negative samples for $v_2$.
In practice, not all nodes are required to participate in the calculation, we can focus on the union of context nodes of $v_1$ and $v_2$.
Finally, the objective function of the pairwise similarity model can be concluded as:
\begin{align}
    \mathcal{J}_{pair} &= -\frac{1}{|\mathcal{P}|} \sum_{x \in \mathcal{P}} \Big(\log{\sigma(f(v) \cdot f'(u))} \nonumber \\
    & \qquad \qquad \quad - \sum_{x_{n} \in \mathcal{P}_{n}}{\log{\sigma(f(v_n) \cdot f'(u_n))}}\Big),
\end{align}
\noindent where $\mathcal{P}$ is the set of positive pairs, 
$x=(v,u)$ is a positive pair where $v$ is the center node of a window and $u$ is a context node in the window,
$\mathcal{P}_n$ is the set of negative pairs
and $x_n=(v_n,u_n)$ is a negative pair,
where $v_n$ is the center node and $u_n$ is a random node.
The negative sampling for pairs follows the negative sampling of Skip-gram \cite{mikolov2013distributed}: for each positive pair, we pick out the center word $v$, and then randomly sample five other nodes to construct five negative samples.

\paragraph{Tuplewise Similarity}
As discussed above, several consecutive nodes in a hyper-path may possess a tuplewise relationship, thus for a target node, we sample two tuples containing it (one in the left-hand side and one in the right-hand side). However, not all consecutive nodes have a tuplewise relationship, for instance, in a $k$-uniform heterogeneous hyper-network, each hyperedge contains $k$ nodes with $k$ different types, so we can filter out tuples that do not satisfy this property. 
A hyperedge can be considered as a sequence of nodes, where the order of the nodes does not carry information. And the 1D Convolutional Neural Networks (1D ConvNet) are a good choice for handling such data. The 1D ConvNet denoted as $cnn(\cdot)$ consists of a 1D convolution layer and a max-pooling layer. Basically, the 1D ConvNet takes the vector representations of the nodes in the sequence as input and outputs a latent vector representation of the sequence. Finally, the latent representation is entered into the fully connected layer with a nonlinear activation function. The tuplewise similarity of a tuple $y = (y_1,y_2,\cdots,y_k)$ can be formulated as follows:
\begin{equation}
s_{tuple}(y) = \delta_2(y) \cdot \sigma(W \boldsymbol{c} +b),
\end{equation}
\noindent where $W$ and $b$ are the weights and bias of the fully connected layer respectively, $\boldsymbol{c} = cnn([f(y_1);f(y_2);\cdots;f(y_k)])$ is the output of the 1D ConvNet, 
and $\delta_2(y)$ is an indicator function which is defined as follows:
\begin{equation}
\delta_2(y) = 
\begin{cases}
1, & \text{if $y$ satisfies the property} \\
0, & \text{otherwise}
\end{cases}.
\end{equation}
Finally, the object function to maximize $\log{s_{tuple}(y)}$ for any tuple $y$ can be concluded as:
\begin{equation}
    \mathcal{J}_{tuple} = -\frac{1}{|\mathcal{T}|} \sum_{y \in \mathcal{T}} \Big(\log{s_{tuple}(y)} - \sum_{y_n \in \mathcal{T}_{n}}{\log{s_{tuple}(y_n)}}\Big),
\end{equation}
\noindent where $\mathcal{T}$ is the set of positive tuples, $y$ is a positive tuple generated from the hyper-path-based random walks,
$\mathcal{T}_n$ is the set of negative tuples and $y_n$ is a negative tuple.
The negative sampling for tuples follows the rule below: for each positive tuple, we randomly select one node in it, and then randomly sample other nodes which satisfy the property of the hyperedge (e.g., the same type) from the node set five times. 
This is to say, all properties of the negative samples are consistent with the positive samples.

Combining the two losses together, the objective function of Hyper-gram is formulated as follows:
\begin{align}
\mathcal{J} &= \mathcal{J}_{pair} + \lambda \mathcal{J}_{tuple} \nonumber \\
&=-\frac{1}{|\mathcal{P}|} \sum_{x \in \mathcal{P}} \Big(\log{\sigma(f(v) \cdot f'(u))} 
- \sum_{x_{n} \in \mathcal{P}_{n}}{\log{\sigma(f(v_{n}) \cdot f'(u_{n}))}}\Big) \nonumber \\
& \quad - \frac{\lambda}{|\mathcal{T}|} \sum_{y \in \mathcal{T}} \Big(\log{s_{tuple}(y)} - \sum_{y_n \in \mathcal{T}_n} \log{s_{tuple}(y_n)\Big)},
\label{eq:objective}
\end{align}
\noindent where $\lambda$ is a hyperparameter to balance the pairwise loss and tuplewise loss.

\paragraph{\textbf{Complexity Analysis.}} For a $3$-uniform hyper-network $\mathcal{G}(\mathcal{V},\mathcal{E})$, the complexity of our model can be divided into three parts: calculating the indecomposable factor, which takes $O(|\mathcal{E}|)$ time; generating random walks, which takes $O(r|\mathcal{V}|)$ time, where $r$ is the number of samples used, and incurs a space complexity of $O(|\mathcal{E}|^2/|\mathcal{V}|)$ to store the interconnections between the neighbors of each node; learning the representations, which incurs a time complexity of $O(dr|\mathcal{V}|)$, where $d$ is the dimension of the low-dimensional vectors. Since hyper-path-based random walks have obvious preferences for nodes with higher path orders, thus we can pre-sample a certain number of nodes to reduce the space complexity when stimulating random walks.

\section{Experiments}
In this section, we conduct experiments to demonstrate the effectiveness of the hyper-path-based random walks and the performance boost due to the well designed Hyper-gram model. Besides, we design several variations of our proposed model to verify the utility of each part of our model.

\begin{table*}[tp!]
    \caption{Statistics of the datasets.}
    \begin{center}
    \begin{tabular}{c|c|c|c|r|r|r|r}
        \toprule
        Datasets & \multicolumn{3}{c|}{Node Types} & \multicolumn{3}{c|}{$|\mathcal{V}|$} & $|\mathcal{E}|$ \\
        \midrule
        Random & a & b & c & 1,000 & 1,000 & 10 & 5,000 \\ 
        \hline
        GPS & user & location & activity & 146 & 70 & 5 & 1,436  \\
        \hline
        MovieLens & user & movie & tag & 1,943 & 5,610 & 8,040 & 47,957 \\
        \hline
        Drugs & user & drug & reaction & 12 & 1,077 & 6,398 & 171,756 \\
        \hline
        Wordnet & head & relation & tail & 40,750 & 18 & 40,752 & 146,442 \\
        \bottomrule
    \end{tabular}
    \label{table:dataset}
    \end{center}
\end{table*}

\begin{table}[ht]
    \caption{Indecomposable factors of the datasets.}
    \begin{center}
    \begin{tabular}{c|c|c|c}
        \toprule
         Datasets & $\xi_a$ & $\xi_b$ & $\xi_c$ \\
        \midrule
        Random & 1.0001 & 0.9998 & 1.0221 \\
        \hline
        GPS & 0.3897 & 0.8594 & 0.09851  \\
        \hline
        MovieLens & 0.0036 & 0.0041 & 0.0040 \\
        \hline
        Drugs & 0.0084 & 0.2074 & 0.2602 \\
        \hline
        Wordnet & 0.2091 & 0.0236 & 0.2097 \\
        \bottomrule
    \end{tabular}
    \label{table:factor}
    \end{center}
\end{table}

\subsection{Datasets}

We use four different types of datasets, including a GPS network, a social network, a medicine network and a word network. The statistics of these datasets are listed in Table \ref{table:dataset}.
\begin{itemize}
    \item \textbf{GPS} \cite{zheng2010collaborative} The GPS dataset records users' trajectories, points out meaningful locations for recommendation, and provides users' activity annotations for locations. We model the information in the dataset as a (user, location, activity) hyper-network.
    \item \textbf{MovieLens}\footnote{\url{https://grouplens.org/datasets/movielens/}} \cite{harper2016movielens} The MovieLens datasets are widely used in education, research, and industry, which describe member activities in the MovieLens movie recommendation system. We model the information in the datasets as a (user, movie, tag) hyper-network. 
    \item \textbf{Drugs} The FDA Adverse Event Reporting System (FAERS) is a database that contains information on adverse event and medication error reports submitted to FDA. We use the data provided in \cite{tu2018structural} and model the information in the dataset as a (user, drug, reaction) hyper-network.
    \item \textbf{Wordnet} The Wordnet \cite{miller1995wordnet} is an online lexical database designed for producing an intuitively usable dictionary and thesaurus. Its entities correspond to word senses and relationships between them. \citeauthor{bordes2013translating} \cite{bordes2013translating} extract a collection of triplets (synset, relation, synset) from the Wordnet, thus we can model the information as a (head entity, relation, tail entity) hyper-network.
\end{itemize}

Besides, we construct a random hyper-network with randomly generated hyperedges.
For these five hyper-networks, we calculate the indecomposable factors defined in Section \ref{sec:indecomposability}, and the results are listed in Table \ref{table:factor}.
From Table \ref{table:factor}, we find that although MovieLens is a heterogeneous hyper-network, the indecomposable factors are very small, which means that we can break down the MovieLens hyper-network into a conventional network, and tackle it with the traditional pairwise methods. This is consistent with our assumptions in Section \ref{sec:intro}, where we think that the (user, movie), (user, tag) and (movie, tag) relationships can exist independently from the retaining element.
We also find that if a hyper-network is randomly generated, the indecomposable factor is equal to 1, which is consistent with the fact that if the hyperedges of a hyper-network are randomly generated, event $A$ and event $B_t$ are irrelevant, thus $P(B_t|A)=P(B_t)$.

\begin{table*}[ht]
    \caption{AUC scores on link prediction.}
    \begin{center}
    \begin{tabular}{c|ccc|ccc|ccc|ccc}
        \toprule
        &\multicolumn{3}{c|}{GPS}&\multicolumn{3}{c|}{MovieLens}&\multicolumn{3}{c|}{Drugs}&\multicolumn{3}{c}{Wordnet} \\
        & \textit{L1}& \textit{L2}& \textit{COS} & \textit{L1}& \textit{L2} & \textit{COS} & \textit{L1}& \textit{L2} & \textit{COS} & \textit{L1}& \textit{L2} & \textit{COS}\\
        \midrule
        DeepWalk & 0.7343 & 0.7413 & 0.8113  & 0.9629 & 0.9643 & 0.9520 &  0.8935 & 0.8994 & 0.7674 & 0.8630 & 0.8675 & 0.8835 \\
        Node2vec & 0.7986 & 0.7996 & 0.8461 & 0.9632 & 0.9645 & 0.9524 & 0.8977 & 0.9013 & 0.7887 & 0.8642 & 0.8684 & 0.8843 \\
        Metapath2vec & 0.7951 & 0.7979 & 0.8291 & 0.8981 & 0.9137 & 0.9387 & 0.8527 & 0.8592 & 0.8198 & 0.8554 & 0.8589 & 0.8612 \\
        LINE     & 0.7573 & 0.7520 & 0.8119 & 0.9117 & 0.8844 & 0.8425  & 0.8828 & 0.8391 & 0.8042 & 0.7740 & 0.7606 & 0.8020 \\
        \hline
        HHE  & 0.6886 & 0.6814 & 0.6843 & 0.7826 & 0.7386 & 0.7325 & - & - & - & - & - & - \\
        HGE   & 0.7405 & 0.7413 & 0.7364 & 0.8912 & 0.8904 & 0.8874 & 0.9269 & 0.9248 & 0.9230 & 0.6454 & 0.6461 & 0.6327 \\
        DHNE  & \multicolumn{3}{c|}{*0.9336} & \multicolumn{3}{c|}{*0.9307} & \multicolumn{3}{c|}{*0.9498} & \multicolumn{3}{c}{*0.7512} \\
        \hline
        \textbf{HPSG} & 0.8799 & 0.8845 & 0.8661 & 0.9628 & \textbf{0.9649} & 0.9530 & 0.9343 & 0.9357 & 0.8846 & 0.8697 & 0.8716 & 0.8740 \\
        \textbf{HPHG} & \multicolumn{3}{c|}{*\textbf{0.9463}} & \multicolumn{3}{c|}{*0.9475} & \multicolumn{3}{c|}{*\textbf{0.9678}} & \multicolumn{3}{c}{*\textbf{0.9012}} \\
        \hline
        \multicolumn{13}{l}{*DHNE and HPHG combine a tuplewise similarity function as the metric.} \\
    \end{tabular}
    \label{table:link_prediction}
    \end{center}
\end{table*}

\subsection{Experimental Settings}
\subsubsection{Baselines}

We compare our models with the following baselines:

\begin{itemize}
    \item Traditional network representation learning methods:
    \begin{itemize}
        \item \textbf{DeepWalk} \cite{perozzi2014deepwalk} DeepWalk uses local information obtained from truncated uniform random walks and learns latent representations by the Skip-gram model \cite{mikolov2013distributed}. To apply this algorithm to hyper-networks, we construct pairwise networks by the clique expansion \cite{sun2008hypergraph}, where any two nodes in the same hyperedge are connected by an edge. And this transformation method is also applied to other traditional baselines.
        \item \textbf{Node2vec} \cite{grover2016node2vec} Node2vec can be considered as an extension of DeepWalk, where a second order random walk approach is utilized to capture the structural information in the conventional networks. 
        \item \textbf{Metapath2vec} \cite{dong2017metapath2vec} Metapath2vec is a meta-path-based representations learning model for heterogeneous information networks. We try all possible meta-path schemes and report the best result among them. Given the GPS datasets with (user (U), location (L), activity (A)) relationships as an example, we try schemes "ULALU", "UALAU", "LUAUL", "LAUAL", "AULUA" and "ALULA".
        \item \textbf{LINE} \cite{tang2015line} LINE introduces an objective function which preserves both first order and second order proximities in the conventional networks.
    \end{itemize}
    \item Hyper-network representation learning methods:
    \begin{itemize}
        \item \textbf{HHE} \cite{zhu2016heterogeneous} HHE is designed to exploit various high-order relationships without information loss. The goal of its cost function is to optimize all relationships in hyper-networks jointly. 
        \item \textbf{HGE} \cite{yu2018modeling} HGE aims to make the representations of nodes contained in a hyperedge close to each other by incorporating multi-way relationships into an optimization problem related to geometric mean and arithmetic mean.
        \item \textbf{DHNE} \cite{tu2018structural} DHNE is a neural network model designed for the link prediction task, which aims to learn a nonlinear tuplewise similarity function for hyperedges and preserve both local and global proximities in the vector space.
    \end{itemize}
\end{itemize}

Based on the proposed methods described in the paper, our models are as follows:

\begin{itemize}
    \item \textbf{HPSG} (\textbf{H}yper-\textbf{p}ath-based random walks + \textbf{S}kip-\textbf{g}ram): This model is a combination of the hyper-path-based random walks (Section \ref{sec:hyper-path}) and the Skip-gram model. This model only captures the pairwise relationships in the random walks, thus we use traditional pairwise metrics to evaluate the pairwise similarity of two nodes. This model is suitable for hyper-networks with small indecomposable factors. 
    \item \textbf{HPHG} (\textbf{H}yper-\textbf{p}ath-based random walks + \textbf{H}yper-\textbf{g}ram): This model is a combination of the hyper-path-based random walks (Section \ref{sec:hyper-path}) and the Hyper-gram model (Section \ref{sec:hyper-gram}). This model captures the pairwise and tuplewise relationships in the random walks at the same time. Besides, this model combines with a nonlinear tuplewise similarity function as the metric for tuplewise relationship evaluation. 
\end{itemize}

\subsubsection{Parameter Settings}

We uniformly set embedding size as 32 for all methods and have a fair parameter tuning for all methods. For algorithms based on random walks like DeepWalk, Node2vec, and Metapath2vec, we set window size as 6, walk length as 80, number of walks as 10. We do grid search over $p, q \in \{0.25, 0.5, 1, 2, 4\}$ for the second order random walk model of Node2vec. We set the number of epochs of Skip-gram as 15 for GPS dataset and 5 for other datasets (GPS is a small dataset that needs more epochs). For LINE, we set the number of negative samples as $5$. For HHE, HGE, and DHNE, we follow the default setting and turn some important parameters such as the number of epochs. For our proposed hyper-path-based random walk model and Hyper-gram model, we set $\alpha$ as 100 for GPS, MovieLens and Drugs dataset, and 20 for Wordnet dataset with a pre-sampling skill. And we set the number of epochs of Hyper-gram as 5 for GPS dataset and 1 for other datasets, and $\lambda$ as 1 for all datasets. Other parameter settings are consistent with random walk-based models like DeepWalk.

\subsection{Link Prediction}

\begin{table}[tp!]
\caption{Pairwise similarity metrics for node pair $(v, u)$ with embeddings $f(v) \in \mathbb{R}^d$ and $f(u) \in \mathbb{R}^d$ respectively. $f_i(\cdot)$ denotes the $i$-th element of $f(\cdot)$.}
\begin{center}
\begin{tabular}{c|c}
\toprule
Metric & Definition \\
\midrule
Weighted-L1 (L1) & $\sum_{i=1}^d |f_i(v)-f_i(u)|$  \\
Weighted-L2 (L2) & $\sqrt{\sum_{i=1}^d (f_i(v)-f_i(u))^2}$ \\
Cosine Similarity (COS) & $\frac{\sum_{i=1}^d f_i(v) f_i(u)}{\sqrt{\sum_{i=1}^d f_i^2(v)} \cdot \sqrt{\sum_{i=1}^d f_i^2(u)}}$ \\
\hline
\end{tabular}
\label{table:metric}
\end{center}
\end{table}

Link prediction aims to predict the potential links of a network, which is a widely-used application in various fields, such as recommender system, link mining and so forth. In this work, link prediction is equivalent to hyperedge prediction that we predict the potential hyperedges of a given hyper-network.

For those traditional models which only consider pairwise relationships, and HHE and HGE which do not train a tuplewise similarity function, there are some binary operations to measure the pairwise relationship of two nodes in Table \ref{table:metric}. 
To measure the tuplewise relationship in hyper-networks, we calculate the mean among all pairwise relationship metric values in a candidate hyperedge. The performance of the three pairwise similarity metrics varies with models and datasets, so in our experiments, we calculate all the three metrics for pairwise similarity-based models.

For all datasets, we hide 20\% of the known hyperedges during training and the hidden hyperedges are used as the positive samples when testing. And the negative test samples are generated by the following rule: for each positive hyperedge $e$, we generate a negative sample which has 90\%  probability of containing one node in $e$ and have 10\% probability of containing two nodes in $e$. Besides, unlike DHNE, our model does not use this prior knowledge (i.e., set the training negative samples to the same distribution), because network representation learning models should not have too much prior knowledge from the test set when training. Each algorithm runs five times and the Area Under Curve (AUC) scores are shown in Table \ref{table:link_prediction}, where we have the following key observations:
\begin{itemize}
    \item For the hyper-networks which have a certain degree of indecomposability (GPS, Drugs, and Wordnet), the models based on the tuplewise similarity function (DHNE and HPHG) outperform the models based on pairwise relationships modeling significantly most of the time. However, for the hyper-network with small indecomposable factors (MovieLens), those pairwise specific models such as DeepWalk, Node2vec, and HPSG perform well.
    \item HPSG outperforms the baselines based on pairwise relationships modeling on different metrics most of the time, which demonstrates that our designed hyper-path-based random walks can well preserve the pairwise relationships of the original hyper-network.
    \item HPHG outperforms DHNE and achieves the best performances on the hyper-networks with relatively high indecomposability. This impressive improvement demonstrates the effectiveness of the Hyper-gram model and well designed hyper-path-based random walk model. 
\end{itemize}

\subsection{Hyper-Network Reconstruction}

\begin{figure*}[ht]
\centerline{\includegraphics[width=0.9\linewidth]{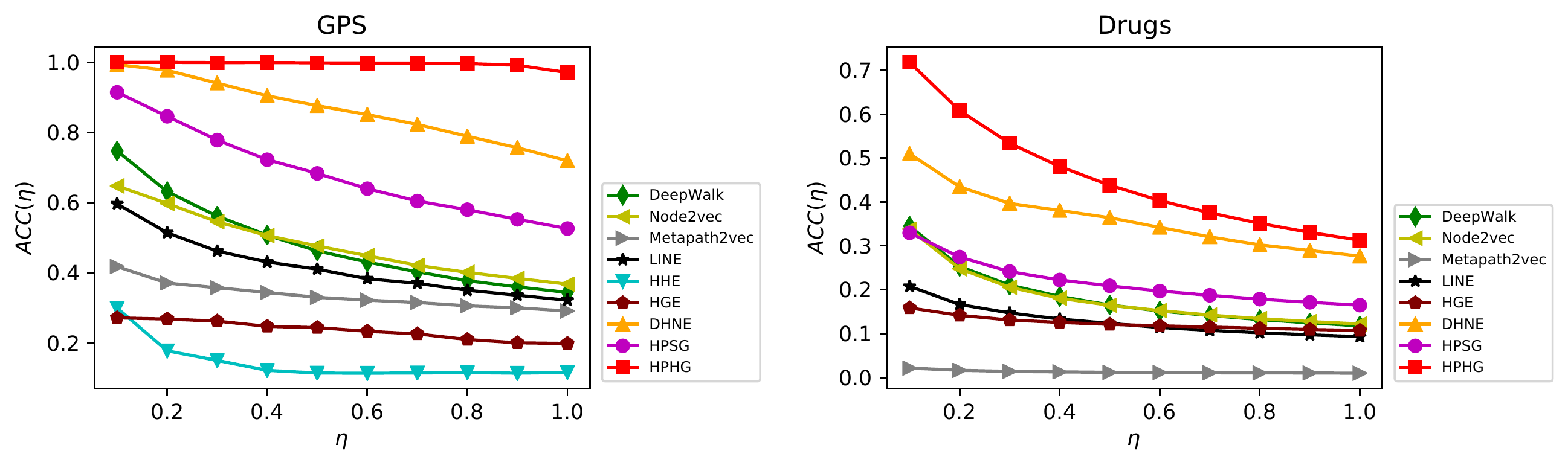}}
\caption{Results of hyper-network reconstruction on GPS and Drugs datasets.}
\label{fig:reconstruction}
\end{figure*}

Good representations of nodes should well preserve the structural information of the original network. And a typical way to evaluate the quality of node representations is to reconstruct the network. Specifically, for a conventional network, we learn the low-dimensional representations of nodes and rank all possible edges according to the pairwise similarities. For hyper-networks, the task of network reconstruction converts to hyper-network reconstruction, where we rank all possible hyperedges according to the pairwise or tuplewise similarities. Compared to traditional network reconstruction, hyper-network reconstruction is a quite difficult task, for the number of possible hyperedges increases exponentially with the max degree of the hyperedges and positive hyperedges like $\{a_1,b_1,c_1\}$ and $\{a_2,b_1,c_1\}$ are really confused for negative hyperedges like $\{a_3,b_1,c_1\}$ (refer to Fig. \ref{fig:hyper-network} to see the hyper-network). Among our datasets, GPS and Drugs are dense $3$-uniform hyper-networks with a relatively small number of possible hyperedges, thus it is possible for us to conduct hyper-network reconstruction on these two datasets.

A common way to measure the performance of network reconstruction is to calculate the percentage of hyperedges in the reconstructed hyper-network that exist in the original hyper-network.
In order to evaluate the performance at different reconstruction percentages, we define the evaluation metric for network reconstruction as follows:
\begin{equation}
ACC(\eta) = \frac{1}{\eta |\mathcal{E}|} \sum_{i=1}^{\eta |\mathcal{E}|} r_i \quad (0 < \eta \leq 1),
\end{equation}
\noindent where $r_i = 1$ means that the $i$-th candidate hyperedge (the hyperedge with the $i$-th largest tuplewise similarity) exists in the original hyper-network and $r_i = 0$ otherwise. $\eta$ is a fraction to control the percentage of hyperedges reconstructed, for instance, $\eta = 0.1$ means that we only reconstruct 10\% of hyperedges, and $\eta = 1$ means that we reconstruct the entire hyper-network. For the models based on the pairwise similarity function, we conduct experiments on all the three metrics for each model and select the best one to report. The results of hyper-network reconstruction are shown in Fig. \ref{fig:reconstruction}.

From the results, we find that our proposed model HPHG outperforms all the baselines significantly, which reconstructs 90\% of hyperedges perfectly and reconstructs the entire hyper-network with an accuracy of 0.9708 on GPS dataset. Besides, our proposed model HPSG beats all the pairwise baselines, which demonstrates that hyper-path-based random walks can well capture the pairwise relationships in hyper-networks. In addition, the models combining with a tuplewise similarity function have an impressive performance boost, which shows the necessity of the modeling of the tuplewise similarity.

In conclusion, the above results demonstrate that our models can better preserve the structural information of hyper-networks by applying the hyper-path-based random walks and the Hyper-gram model which combines both pairwise and tuplewise similarities.

\subsection{Ablation Study}

In this section, we will separate HPHG to verify the utility of the submodels. We designed several variations of HPHG and conduct experiments in link prediction on GPS dataset. The variations are designed as follows:

\begin{itemize}
    \item HPSG: This variation is a combination of the hyper-path-based random walks and the Skip-gram model, where we do not capture the tuplewise similarity in the generated random walks.
    \item HPHG ($\text{window}=0$): This variation is equivalent to the combination of the hyper-path-based random walks and the tuplewise similarity model, where we do not capture the pairwise relationships in the random walks.
    \item HPHG ($\alpha = 0$): This variation is equivalent to the combination of the traditional random walks and the Hyper-gram model.
    \item HPHG: This model is a combination of the hyper-path-based random walks and the Hyper-gram model.
\end{itemize}

\begin{table}[tp!]
\caption{AUC scores of link prediction on GPS dataset for the variations of our model.}
\begin{center}
\begin{tabular}{c|c}
\toprule
Variations & AUC score \\
\midrule
HPSG & 0.8845 \\
HPHG ($\text{window}=0$) & 0.9253 \\
HPHG ($\alpha = 0$) & 0.9060  \\
\hline
HPHG & 0.9463 \\
\hline
\end{tabular}
\label{table:submodel}
\end{center}
\end{table}

The results listed in Table \ref{table:submodel} demonstrate that for the hyper-network with a high degree of indecomposability, both pairwise similarity and tuplewise similarity are necessary to preserve the structure of the original hyper-network. When we replace the Hyper-gram model with Skip-gram model, the tuplewise relationships in the random walks are not captured and the tuplewise similarity function is replaced by the traditional pairwise similarity function. When we set $\text{window} = 0$, there are no pairs captured by the model, and the representations of nodes are only trained by the tuplewise similarity model. When we set $\alpha = 0$, the hyper-path-based random walks decay to traditional random walks, but benefiting from the filter layer, the model can also capture several tuples to train the tuplewise similarity function. In conclusion, the results well verify the validity of each part of our model and demonstrate the relevance and complementarity of our submodels.

\section{Conclusion}

In this paper, we firstly investigate that different hyper-networks have different degrees of indecomposability, and then design the indecomposable factor to evaluate the indecomposability of hyper-network.
Secondly, we propose the concept of hyper-path which well implies the pairwise and tuplewise relationships of the nodes in hyper-networks. Having a good assessment of the indecomposability of hyper-network based on the indecomposable factor, we design the hyper-path-based random walk model to generates random walks which well preserve the structural information of hyper-networks. Finally, we propose a deep model called Hyper-gram which combines both pairwise and tuplewise similarities contained in the generated random walks meanwhile training a nonlinear tuplewise similarity function to evaluate the tuplewise relationships. We conduct extensive experiments on link prediction and hyper-network reconstruction tasks, experimental results demonstrate that our proposed methods can well discover those unobserved links and reconstruct the hyper-networks with a minimal error compared with the state-of-the-art methods. 
Indeed, there are more possible applications for the hyper-path-based random walks and the Hyper-gram model. For instance, we can conduct search and ranking tasks on hyper-networks via hyper-path-based random walks or mine the tuplewise relationships for other kinds of data, e.g., knowledge graph, based on the Hyper-gram model.

\begin{acks}
This paper was supported by the Early Career Scheme (ECS, No. 26206717) from Research Grants Council in Hong Kong.
\end{acks}

\bibliographystyle{ACM-Reference-Format}
\balance
\bibliography{HPHG}

\end{document}